\documentclass[sn-mathphys,Numbered]{sn-jnl}

\usepackage{graphicx}%
\usepackage{multirow}%
\usepackage{amsmath,amssymb,amsfonts}%
\usepackage{amsthm}%
\usepackage{mathrsfs}%
\usepackage[title]{appendix}%
\usepackage{xcolor}%
\usepackage{textcomp}%
\usepackage{manyfoot}%
\usepackage{booktabs}%
\usepackage{algorithm}%
\usepackage{algorithmicx}%
\usepackage{algpseudocode}%
\usepackage{listings}%
\usepackage{soul}

\theoremstyle{thmstyleone}%
\theoremstyle{thmstyletwo}%

\theoremstyle{thmstylethree}%

\begin{document}

\title[Water Electrofreezing]{Electrofreezing of Liquid Water at Ambient Conditions}


\author*[1]{\fnm{Giuseppe} \sur{Cassone}}\email{cassone@ipcf.cnr.it}

\author*[2,3]{\fnm{Fausto} \sur{Martelli}}\email{fausto.martelli@ibm.com}


\affil[1]{\orgdiv{Institute for Chemical-Physical Processes}, \orgname{National Research Council}, \orgaddress{\street{Viale F. Stagno d'Alcontres 37}, \city{Messina}, \postcode{98158}, \country{Italy}}}

\affil[2]{\orgname{IBM Research Europe}, \orgaddress{\street{Keckwik Lane}, \city{Daresbury}, \postcode{WA4 4AD}, \country{United Kingdom}}}

\affil[3]{\orgdiv{Department of Chemical Engineering}, \orgname{University of Manchester}, \orgaddress{\street{Oxford Road}, \city{Manchester}, \postcode{M13 9PL}, \country{United Kingdom}}}


\abstract{
Water is routinely exposed to external electric fields (EFs). Whether, e.g., at physiological conditions, in contact with biological systems, or at the interface of polar surfaces in countless technological and industrial settings, water responds to EFs on the order of a few V/{\AA} in a manner that is still under intense investigation. Dating back to the $19^{th}$ century, the possibility of solidifying water upon applying an EF instead of adjusting temperature and pressure -- a process known as electrofreezing -- is an alluring promise that has canalized major efforts since, with uncertain outcomes. In this work, we perform long \emph{ab initio} molecular dynamics simulations \textcolor{black}{of water at ambient conditions exposed at EFs of different intensities. While the response of single water molecules is almost instantaneous, the cooperativity of the hydrogen bonds induces slower reorganizations that can be captured by dividing the trajectories in disjoint time windows and by performing analysis on each of them separately. Upon adopting this approach, we find} that EFs of $0.10\leq$EFs$\leq0.15$~V/{\AA} induce electrofreezing \textcolor{black}{occurring after $\sim150$~ps. We observe a continuous transition to a disordered state characterized by frozen dynamical properties, damped oscillations, lower energy, and enhanced local structural properties. Therefore, we ascribe this state to} a new ferroelectric amorphous phase, which we term f-GW (ferroelectric glassy water). 
Our work represents the first evidence of electrofreezing of liquid water at ambient conditions and therefore impacts several fields, from \textcolor{black}{fundamental chemical physics to} biology \textcolor{black}{and} catalysis.
}





\maketitle

\section{Introduction}\label{sec1}

With at least 20 known crystalline forms and counting, the baroque phase diagram of water is the most complex of any pure substance~\cite{salzmann2019advances} and is continuously under construction. Two amorphous ices, a low-density amorphous (LDA) and a high-density amorphous (HDA) ice~\cite{mishima1985apparently}, encompass a large set of sub-classes~\cite{amann2016x}; a third, medium density amorphous ice has recently been proposed~\cite{rosu2023medium}, while a plastic amorphous ice has been suggested to exist at high pressures~\cite{zimon2023molecular}. Water is also routinely exposed to external electric fields (EFs). The range of strengths $0.1-1$~V/{\AA} is particularly relevant, as it represents the range continuously produced by molecular dipoles fluctuations~\cite{Geissler} in aqueous solutions~\cite{Chalmet_JCP2001,Saykally_PNAS2005,Ruiz-Lopez_PNAS2021}
and to which water is exposed in countless technological/industrial settings~\cite{che}.
Recent developments have shown that the reaction rates of common organic reactions can be increased by one to six orders of magnitude upon applying external EFs~\cite{Shaik2016,Cassone_ChemSci,aragones,Huang_SciAdv2019,shaik2,Head_Gordon_Natcommun2022}, hence paving the way to the adoption of EFs as efficient catalyzers. Comparable EFs, generated by charge separation, endow microdroplets with strong and surprising catalytic power~\cite{Zare_JACS2019,Zare_JPCL2020,Zare_PNAS2023,martins2023electrostatics}. 

Historically, the possibility of manipulating water kinetics via EFs was first proposed by Dufour in 1862~\cite{Dufour_1862}. As experimental techniques matured over the years, such an opportunity became more tangible: the role of EFs on the heterogeneous nucleation of ice in cirrus clouds was addressed in the 1960s~\cite{Pruppacher1963}, and several other investigations followed, starting a vivid scientific debate~\cite{Doolittle,Sivanesan,Wei,Orlowska,Peleg,Acharaya}. Recently, Ehre et al.~\cite{Ehre} have shown that the kinetics of electrofreezing of supercooled water on pyroelectric materials is highly heterogeneous, favoring the crystallization on positively charged surfaces. \newline
Early -- and pioneering -- computational investigations based on classical molecular dynamics simulations \textcolor{black}{also joined forces. According to these studies, liquid water undergoes electrofreezing to crystalline ice when exposed to external static EFs in the order of $\sim0.5$~V/{\AA}~\cite{Kusalik_PRL,Kusalik_JACS}, and the effects of oscillating EFs have also been investigated~\cite{English_MolPhys}}. On the other hand, \emph{ab initio} molecular dynamics (AIMD), which account for chemical reactions, and experiments have more recently shown that $\sim0.3$~V/{\AA} represents a threshold above which water molecules undergo dissociation into oxonium (H$_3$O)$^+$ and hydroxide (OH)$^-$ ions~\cite{Saitta_PRL,Cassone_JPCL,Stuve2012,Hammadi2012}. Seemingly, below this threshold, thermal energy and the associated large intrinsic field fluctuations taking place at the molecular scale impede the ordering of the hydrogen bond network (HBN), a necessary step for crystallization to occur. This task can instead be achieved, according to classical simulations, upon tuning the working pressure to $\sim5$~kPa and imposing external EFs of $\sim0.2$~V/{\AA}~\cite{Zhu2019}. 

The application of EFs to liquid water \textcolor{black}{induces a fast response of water molecules which align their dipole parallel to the field direction. On the other hand, the intrinsic cooperativity of HBs acts as a competing force, slowing down the relaxation and, in turn,} driving the sample out of equilibrium. Therefore, in order to follow the response of water, it is necessary to probe the system at (non-)overlapping time windows rather than averaging over the entire simulation, \textcolor{black}{and this is the paradigm we  have decided to adopt (we report, in the SI, a comparison between quantities of interest computed at disjoint time windows and averaged over entire trajectories)}. In this study, we perform long ($\sim250$~ps) AIMD simulations and show that EFs in the order of $0.10\leq$EFs$\leq0.15$~V/{\AA} induce a structural transition to a new ferroelectric glassy state that we will call f-GW (ferroelectric glassy water). This transition occurs after $\sim150$~ps and is signaled by the freezing of the translational degrees of freedom, the suppression of the fluctuations of the HBN, and the drop in the potential energy. Our work represents the first evidence of electrofreezing of liquid water occurring at \emph{ambient conditions}.

\section{Results}\label{sec2}
In Fig.~1 we report the infrared (IR) spectra for bulk water without field (violet line) and for increasingly higher applied fields ($0.05$~V/{\AA}, blue; $0.10$~V/{\AA}, orange; $0.15$~V/{\AA}, red) computed over the \textcolor{black}{time window $[200-250]$}~ps of the respective trajectories. In the absence of applied fields, the position of the OH stretching band -- located at $3220$~cm$^{-1}$ -- and that of low-frequency libration mode -- at $560$~cm$^{-1}$ -- are in good agreement with the experimental data~\cite{Bertie}. Upon exposure of the water sample to external EFs, we observe a contraction of the frequency range ascribed to the vibrational Stark effect~\cite{vibStark2}, as also reported in Ref.~\cite{Cassone_PCCP19} \textcolor{black}{and -- on \textcolor{black}{limited frequency domains} -- in Ref.~\cite{Futera_English17}}. This contraction indicates that the field imposes novel selection rules on the molecular vibrations. The largest frequency shift is associated with the OH stretching band; the corresponding red-shift is in the order of $\sim75$~cm$^{-1}$ each $0.05$~V/{\AA}, up to an EF of $0.10$~V/{\AA}. However, a milder further red-shift in the order of $35$~cm$^{-1}$ occurs at a field of $0.15$~V/{\AA}. The red-shift of the OH stretching is generally associated with stronger hydrogen bonds (HBs)~\cite{rice} and the development of more ``ice-like'' environments~\cite{wang_jpca}. The reduced magnitude of the relative red-shift upon increasing the field from $0.10$~V/{\AA} to $0.15$~V/{\AA}, as quantified by the difference in frequencies reported in the inset of Fig.~1, suggests that the effect of the applied field becomes less intense.
Moving towards lower frequencies, the weak libration+bending combination mode band at $2200$~cm$^{-1}$ is commonly associated with the strength of the hydrogen bond network (HBN)~\cite{verma}. The presence of the external EFs induces an enhancement and a concurrent slight blue-shift of this band, further suggesting that the EF causes a strengthening of the HBN. A similar effect has been reported on the IR spectra of water undergoing supercooling~\cite{perakis2011two} \textcolor{black}{where the strengthening of the HBN is, instead, induced by the reduction of thermal energy}.
A stronger effect on the vibrational spectrum occurs at lower frequencies, the signature of librational modes. The application of an external EF induces a significant blue-shift and, at $0.10$~V/{\AA} and $0.15$~V/{\AA}, the development of a clear new band peaked at $\sim1000$~cm$^{-1}$. This band has been ascribed to the breaking of the isotropy of molecular rotations and the preferential alignment with the field direction~\cite{Cassone_PCCP19}, as shown in Fig.~S1 of the SI.

The picture emerging from the inspection of the IR spectra, therefore, indicates that the EF affects the topology of the HBN in several ways: the red-shift of the OH stretching band occurring upon increasing the applied EF is indicative of a strengthening of the HBs, while the blue-shift of the libration+bending combination mode band and that of the librations at lower frequencies suggests some degree of ordering of the HBN. At the same time, the appearance of a new peak in the librational band indicates an alignment of the molecular dipoles along the field direction.

The strengthening of the HBs \textcolor{black}{(or their stiffening, as shown in Fig. S7)}  and the alignment of the molecular dipoles along the field direction mirror an enhancement of spatial correlations that also persists over time. In order to test this hypothesis, we report, in Fig.~2, the $G_{OO}(r,t)$, the Van Hove correlation function computed between oxygen atoms only \textcolor{black}{and in the same time window $[201-250]$~ps on which we have computed the IR spectra shown in Fig.~1}. In Fig.~2 a) we report $G_{OO}(r,t)$ in the absence of external fields. We can observe that weak spatial correlation in the region $\sim2.8$~{\AA} and $\sim4.5$~{\AA}, corresponding to the first and second shells of neighbours, rapidly \textcolor{black}{wear} off in timescales of $\sim5-10$~ps. The application of a field of $0.05$~V/{\AA}, reported in panel b), induces an extension of spatial correlations over slightly longer timescales. Radically stronger responses are induced by more intense fields: a field of $0.10$~V/{\AA} (panel c)) and a field of $0.15$~V/{\AA} (panel d)) clearly strengthen spatial correlations between $\sim2-3$~{\AA} and $4-5$~{\AA} and extend them to timescales above $\sim35$~ps. \newline

\begin{figure}[h!]
    \centering
    \includegraphics[width=\textwidth]{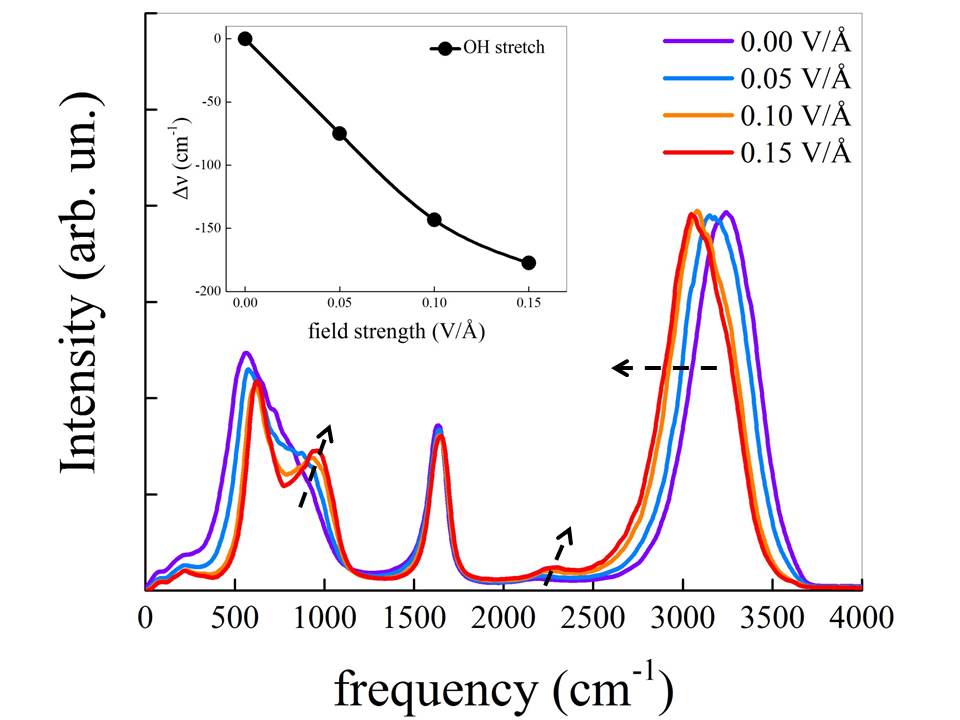}
    \caption{Infrared (IR) absorption spectra of liquid water determined at zero field (violet line) and under different field intensities as detailed in the legend. Arrows are guides for the eye qualitatively following the field-induced modifications of the bands. In the inset, we report the vibrational Stark effect of the OH stretching band. \textcolor{black}{Data are computed on the time window $[201-250]$~ps.}}
    \label{fig:fig1}
\end{figure}
\begin{figure}[h!]
    \centering
    \includegraphics[width=\textwidth]{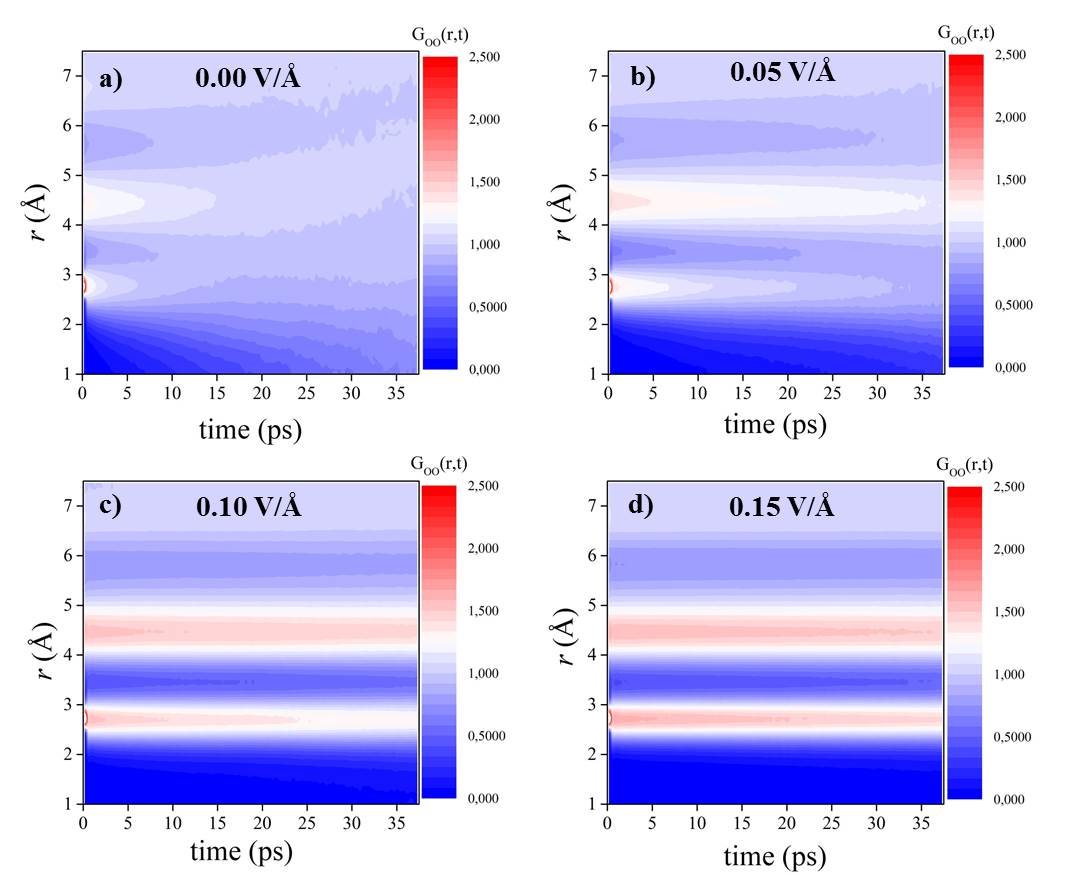}
    \caption{Partial Van Hove correlation functions between the oxygen atoms (i.e., $G_{OO}(r,t)$) as a function of the time and of the intermolecular distance in the absence of the field (a) and in presence of static electric fields with intensities equal to $0.05$ (b), $0.10$ (c), and $0.15$~V/{\AA} (d). \textcolor{black}{Data are computed on the time window $[201-250]$~ps.}}
    \label{fig:fig1}
\end{figure}
\begin{figure}[h!]
    \centering
    \includegraphics[width=\textwidth]{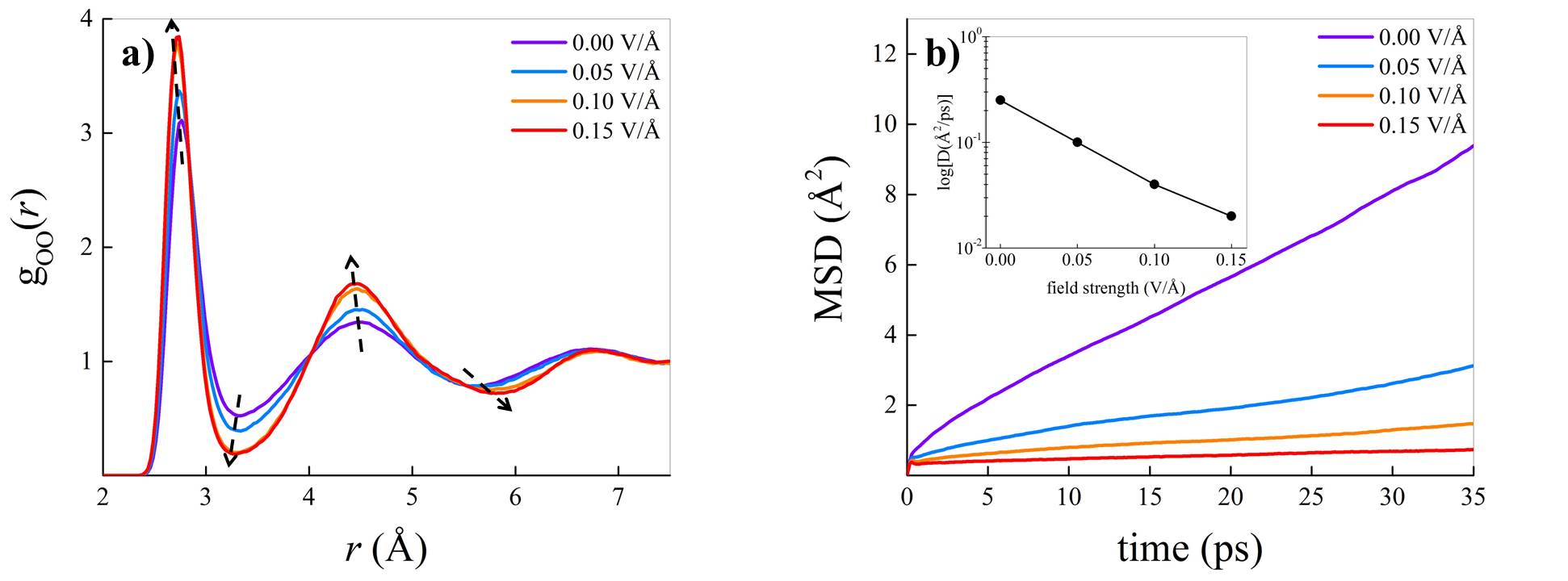}
    \caption{(a) Oxygen-oxygen radial distribution functions at different electric field strengths (see legend). Dashed arrows qualitatively depict field-induced modulation of the hydration shells. (b) Mean squared displacement (MSD) of the oxygen atoms at various field intensities (see legend). In the inset, a logarithmic plot of the self-diffusion coefficient of the oxygen atoms as a function of the field strength. \textcolor{black}{Data are computed on the time window $[201-250]$~ps.}}
    \label{fig:rdfs}
\end{figure}
By projecting the partial Van Hove correlation functions on the reduced domain constituted by the spatial distances only (i.e., by removing the temporal dependence), we obtain the oxygen-oxygen radial distribution functions $g_{OO}(r)$. In Fig.~\ref{fig:rdfs} a) we report the $g_{OO}(r)$ computed in the time window $[201-250]$~ps. Without any applied field (violet), the $g_{OO}(r)$ is that of bulk liquid water with a first peak located at $\sim2.8$~{\AA} and a second peak at $\sim4.5$~{\AA}. Adding a small EF of $0.05$~V/{\AA} (blue) we observe an increase in the intensity of both the first and the second peaks with a reduction of the population between the first and second peaks. Upon doubling the intensity of the field and reaching $0.10$~V/{\AA} (orange) we observe an enhanced increase in the intensity of both the first and second peaks and a further depletion of water molecules populating the interstitial region. An additional increase in the field intensity to $0.15$~V/{\AA} (red) does not show appreciable changes in the $g_{OO}(r)$ with respect to the previous case, suggesting that no further major structural changes occur in the sample. Fig.~S\textcolor{black}{4} of the SI reports the $g_{OO}(r)$ computed in consecutive time windows of $50$~ps starting from the beginning of our simulations, hence providing a glimpse of the dynamical structural transformations. In agreement with the profiles of the Van Hove functions, it is possible to observe that the $g_{OO}(r)$ for $0.05$~V/{\AA} converges to the same profile after $50$~ps (Fig.~S3-a), while convergence is achieved only after $150$~ps for $0.10$~V/{\AA} (S3-b) and $200$~ps for $0.15$~V/{\AA} (S3-c). We notice, at this point, that the $g_{OO}(r)$ at fields of $0.10$~V/{\AA} and $0.15$~V/{\AA} at convergence, i.e., after $200$~ps of simulation, strikingly resemble the $g_{OO}(r)$ of supercooled water or that of low-density amorphous (LDA)
ice~\cite{amann2016x}. 
\textcolor{black}{This comparison is, instead, less accurate if one does not take into account the out-of-equilibrium nature that drives the process, and computes the radial distribution functions over the entire trajectories, as reported in Fig.~S10-a as well as in several previous works.}
In order to rule out the effect of the simulation box, we have performed longer simulations (up to $\sim500$~ps) for systems with $256$ H$_2$O molecules at densities of $0.92$~g/cm$^3$ and $0.95$~g/cm$^{3}$. Our results, reported in Fig.~S\textcolor{black}{8} of the SI, show that the development of a glassy-like $g_{OO}(r)$ is independent of the system size and density. \newline 
Considering the high computational cost of performing AIMD simulations, we can not produce an equilibrated supercooled sample or an LDA via realistic quenching rates \textcolor{black}{to compare the relative radial distribution functions}. Therefore, in order to understand whether our $g_{OO}(r)$s belong to a glassy sample or to a supercooled sample, we look at dynamical properties, namely the diffusivity measured via the mean squared displacement (MSD). Our results are reported in Fig.~\ref{fig:rdfs} b). We stress here that, like for the $g_{OO}(r)$, the MSD are computed on the time window $[201-250]$~ps. It is possible to appreciate how the slope of the MSD drastically drops as soon as we introduce an EF. In the presence of a weak field of $0.05$~V/{\AA} (blue) the sample is still liquid, although the mobility is strongly reduced compared to the case without field (violet). Upon increasing the field to $0.10$~V/{\AA} (orange) and to $0.15$~V/{\AA} (red) the MSD profiles indicate that water's translational degrees of freedom are confined to molecular vibration and to the rattling within the cage of the local neighbourhood. Computing the MSD over wider time windows implies accounting for the contribution of water molecules still in the liquid phase, hence artificially increasing the slope of the MSD, \textcolor{black}{as shown in Fig.~S10-b}. We posit that this might be one of the reasons why the f-GW phase has been overlooked in previous studies. 

\begin{figure}[h!]
    \centering
    \includegraphics[width=\textwidth]{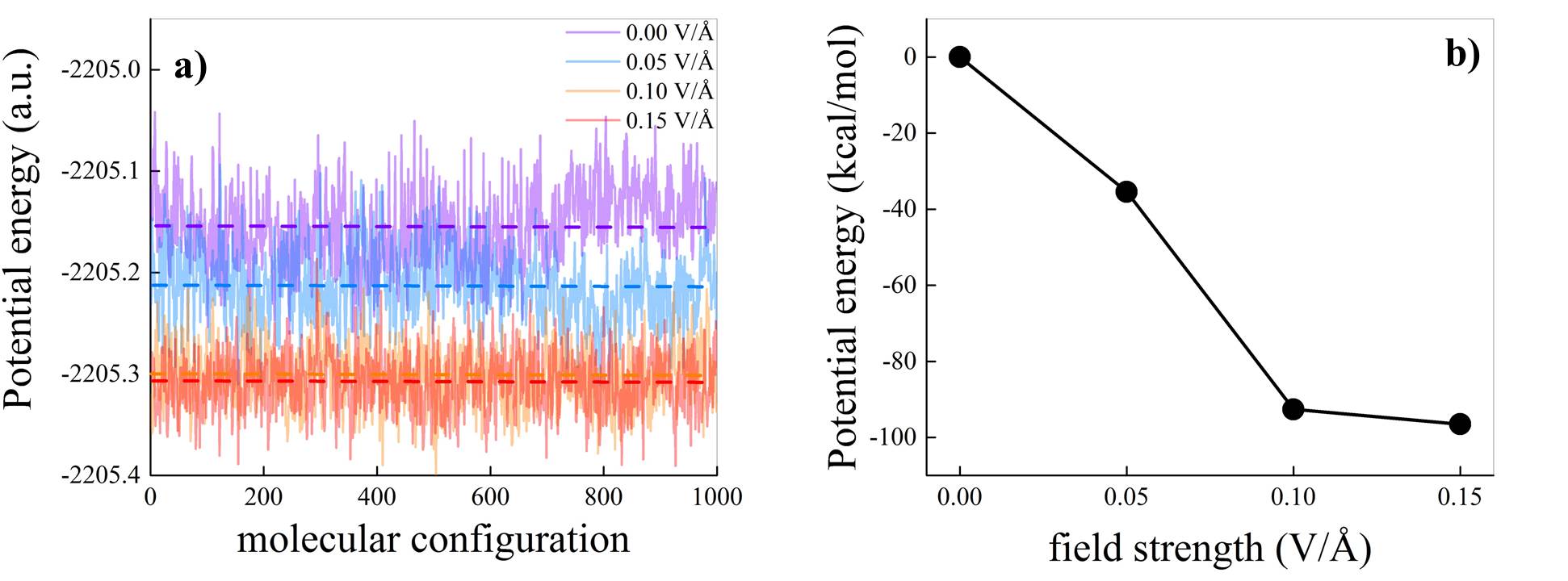}
    \caption{Potential energy computed via single point calculations on $1000$ configurations randomly chosen within the time window $[201-250]$~ps of each simulation. (a) Profile of the potential energy in a.u. for water without field (violet), water in the presence of $0.05$~V/{\AA} (blue), water in the presence of $0.10$~V/{\AA} (orange), and water in the presence of $0.15$~V/{\AA} (red). Dashed lines represent the average value. (b) Average potential energy relative to the zero-field case in kcal/mol as a function of the field strength.}
    \label{fig:pot}
\end{figure}

In Fig.~\ref{fig:pot} we report the profile of the potential energy computed performing single point calculations on $1000$ configurations randomly chosen within the time window $[201-250]$~ps. Panel a) reports the profile as a function of the chosen molecular configurations. Without any applied field (violet) the potential energy fluctuates around the dashed violet line. Upon introducing a field of $0.05$~V/{\AA} (blue) we observe a decrease in potential energy for almost all configurations, with an average value (dashed blue line) sitting below the case of water without field. Stronger drops in potential energy occur in the presence of EFs of $0.10$~V/{\AA} (orange) and of $0.15$~V/{\AA} (red). In panel b) we report the average potential energy -- relative to the zero-field case in kcal/mol -- as a function of the field strength. The drop in potential energy is clearly visible and shows how EFs of $0.10$~V/{\AA} and $0.15$~V/{\AA} drag the system into lower potential energy basins~\cite{debenedetti2001supercooled,sastry2001relationship}. \textcolor{black}{It is worth noticing that the reduction in potential energy occurs along with a reduction of $\sim14\%$ of the entropy, as reported in Ref.~\cite{ContiNibali_JCP23} for the same system and numerical setups.}

\begin{figure}[h!]
    \centering
    \includegraphics[width=\textwidth]{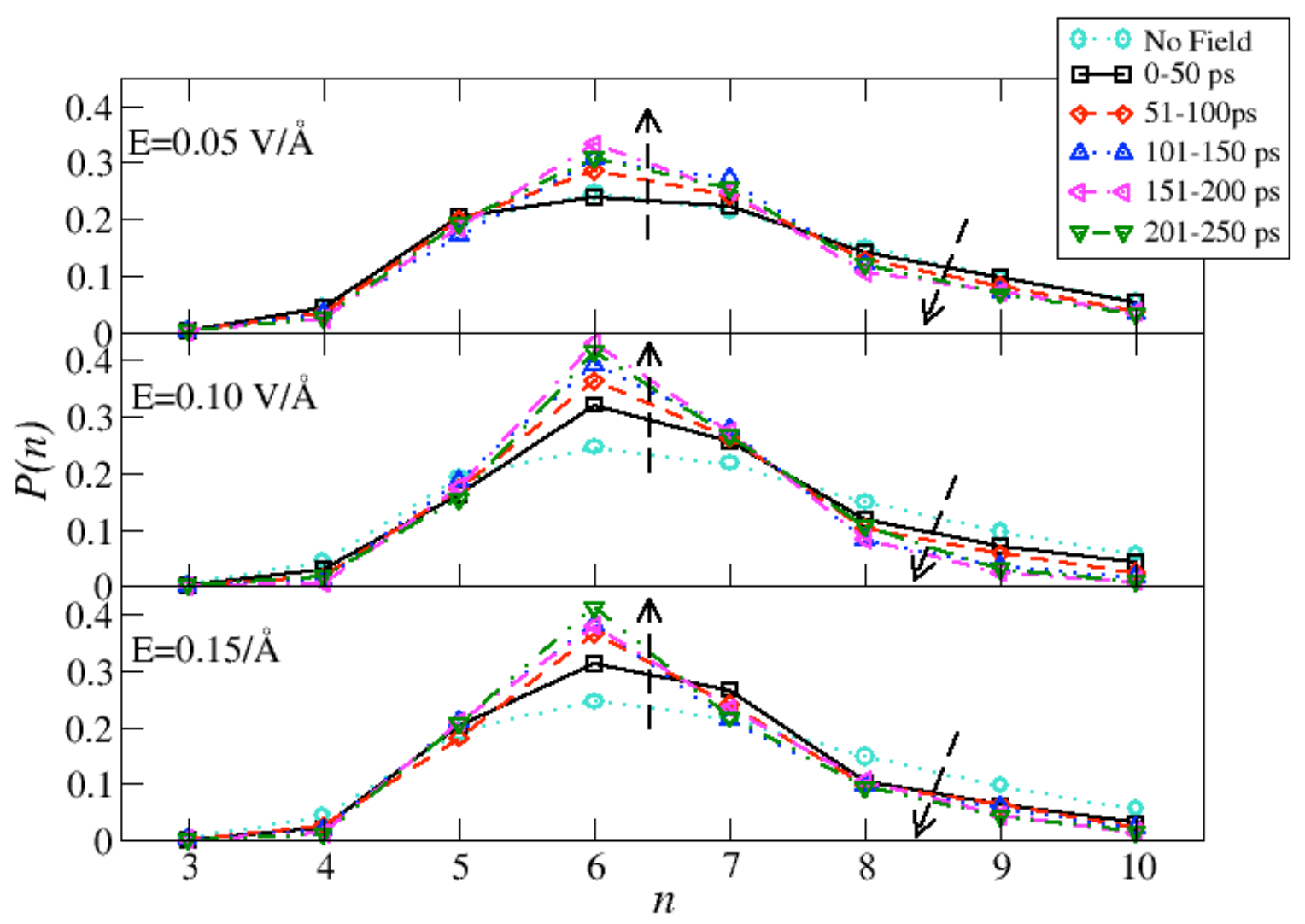}
    \caption{Probability distribution $P(n)$ of having a ring of length $n\in[3,10]$ computed at different time windows during our simulations. The upper panel refers to the applied field $E=0.05$~V/\AA, the middle panel to the applied field $E=0.10$~V/\AA, the lower panel to the applied field $E=0.15$~V/\AA. The cyan circles refer to the zero-field case. The black squares refer to the first $50$~ps, red diamonds to the time window $51-100$~ps, blue upper triangles to the time window $101-150$~ps, the left magenta triangles to the time window $151-200$~ps, and the green lower triangles to the window $201-250$~ps. The dashed arrows emphasize the change in $P(n)$ at consecutive time windows.}
    \label{fig:rings}
\end{figure}

The amorphization of liquid water involves a sensitive change in the fluctuations and topology of the HBN~\cite{martelli2022steady}, which can be quantitatively inspected via the ring statistics. Therefore, in order to confirm that the structural and dynamical changes induced by the EFs indeed prompt a rearrangement in the HBN, we compute $P(n)$, the normalized probability of having a ring of length $n\in[3,10]$ in time windows of $50$~ps. In hexagonal/cubic ice at $0$~K and without defects, the $P(n)$ is centered at $n=6$, indicating that only hexagons are present. In Fig.~\ref{fig:rings} we report $P(n)$ for strengths of $0.05$~V/{\AA}, $0.10$~V/{\AA}, and $0.15$~V/{\AA}. Each case is reported against the $P(n)$ determined in the absence of the EF (cyan circles). In the case of $0.05$~V/{\AA} during the first $50$~ps (black circles, panel a)), we can observe that the topology of the HBN overlaps almost perfectly with that of liquid water. Upon increasing the simulation time, the topology of the HBN responds to the presence of the field by increasing the number of hexagonal and heptagonal rings while reducing the number of longer rings. Overall, the response of the HBN to the presence of a weak field resembles the transformation of the HBN topology upon cooling~\cite{formanek2020probing,martelli2022steady}. \newline 
Upon doubling the field intensity to $0.10$~V/{\AA} (panel b)), the topology of the HBN drastically changes even within the first $50$~ps of simulation. In particular, we observe an increase in hexagonal and heptagonal rings with a corresponding decrease in longer rings. At consecutive simulation time windows, we observe a further sharpening of the $P(n)$ with a considerable increase of hexagonal rings and a depletion of octagonal and longer rings. The topology of the HBN within the last $50$~ps of our simulation is remarkably similar to that of LDA (obtained from classical simulations~\cite{martelli2022steady}).\newline
A similar behaviour occurs when we apply a field of $0.15$~V/{\AA} (panel c)): the HBN reacts to the presence of the field increasing the population of hexagonal and heptagonal rings while decreasing the population of longer rings. Upon increasing the simulation time, the topology of the HBN further increases the population of hexagonal rings while decreasing longer rings, including heptagonal rings.

\begin{figure}[h!]
    \centering
    \includegraphics[width=\textwidth]{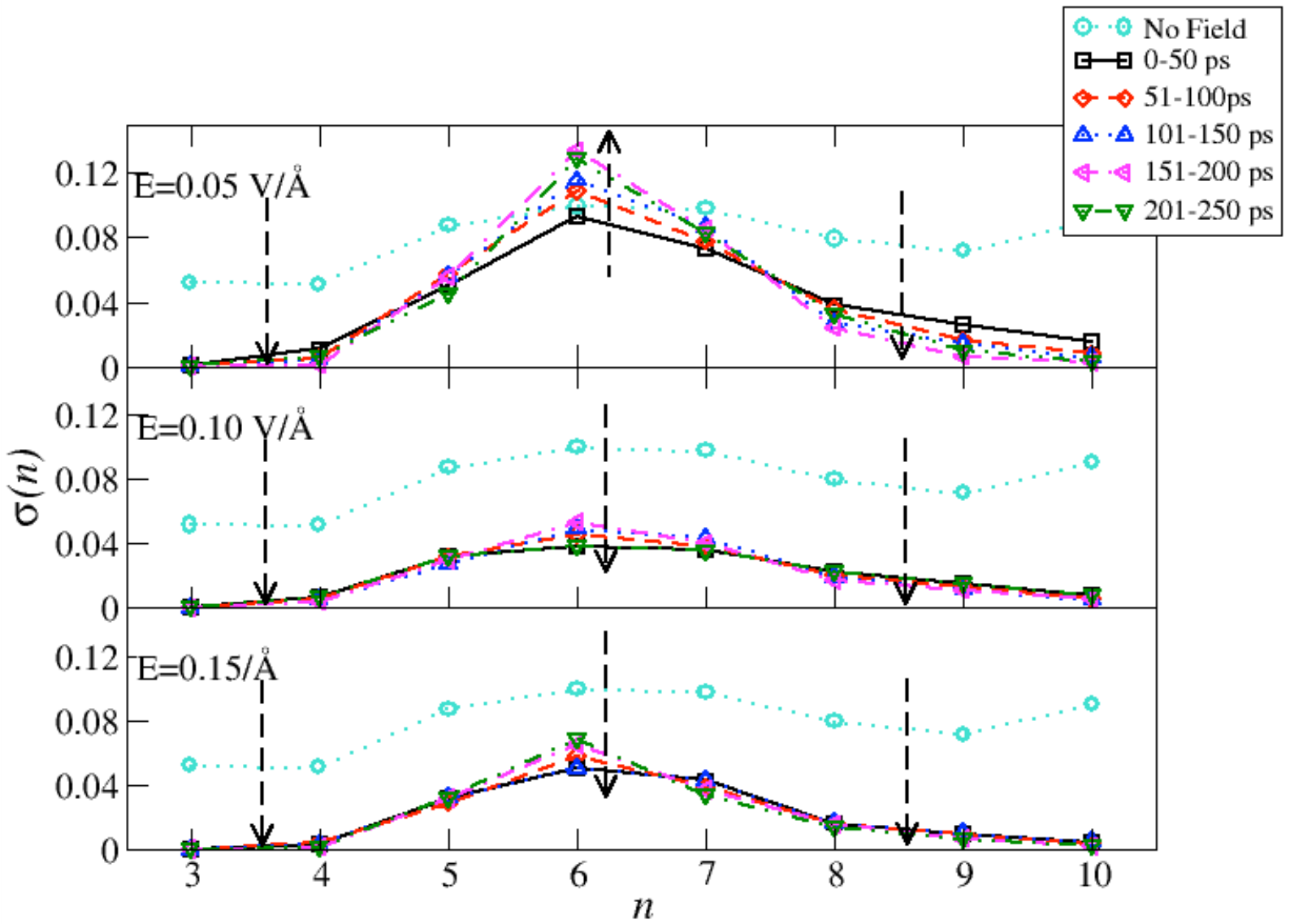}
    \caption{Fluctuations $\sigma(n)$ computed on the ring statistics at different time windows during our simulations. The upper panel refers to the applied field of $E=0.05$~V/\AA, the middle panel to the applied field of $V=0.10$~V/\AA, and the lower panel to the applied field of $E=0.15$~V/\AA. The cyan circles refer to the case of no field. The black squares refer to the first $50$~ps, red diamonds to the time window $51-100$~ps, blue upper triangles to the time window $101-150$~ps, the left magenta triangles to the time window $151-200$~ps, and the green lower triangles to the window $201-250$~ps. The dashed arrows emphasize the $P(n)$ change at consecutive time windows.}
    \label{fig:fluctuations}
\end{figure}

\textcolor{black}{The gradual rearrangement of the topology of the HBN described above occurs on slower timescales compared to the alignment of water's dipole moment (see Fig.~S3) and clearly shows that, although single water molecules react very quickly to the presence of EFs, the overall network of bonds reorganizes itself into new steady configurations on longer times, as also partially reported in Ref.~\cite{jung1999effect}. Such time-dependence, key in our investigation, can be seen in the gradual build-up of four-coordinated water molecules shown in Fig. S\textcolor{black}{9}. This gradual build-up in time leads to an increase in four-coordinated molecules up to $15\%$ from the early stages of the simulation. Such an increase in the percentage of four-coordinated environments also induces a gradual enhancement of the local order. We report, in Fig.~S6, $P(I)$, the probability distribution of the local structure index $I$ estimated on consecutive windows of $50$~ps. It is possible to appreciate the development of bimodality in the later stages of our simulations for fields of $0.10$~V/{\AA} (middle panel) and $0.15$~V/{\AA} (lower panel). The lower panel of Fig.~S6 reports a comparison between $P(I)$ computed in the time window $[201-250]$~ps for $0.15$~V/{\AA} and for LDA at $T=200$~K obtained from classical molecular dynamics simulations. Despite the differences in simulation techniques, the local structure of liquid water under EF strongly resembles that of LDA.}\newline
\textcolor{black}{The information collected so far indicates that our samples gradually readjust to the presence of external EFs. The slow evolution in time involves (i) the gradual development of four-folded configurations interacting via stronger HBs, (ii) the congruent development of more ordered local environments, (iii) the slow reduction of translational and rotational degrees of freedom, (iv) a drop in the potential energy, and (v) the gradual rearrangement of the HBN topology towards configurations richer in hexagonal rings. Eventually, after exposing the samples of liquid water to a field of $0.15$~V/{\AA} for $\sim150$~ps, we observe a complete freezing of translational degrees of freedom, hence suggesting that our sample might be glass. Although the definition of glassy water is precise (molecular relaxation time exceeding $100$~s or the shear viscosity reaching to $1013$ poise), our simulations are too short to access these quantities. On the other hand, it has been recently shown that the transition to glass upon quenching liquid water is clearly signaled by the damping in the fluctuations of the HBN topology~\cite{martelli2022steady}, which we here evaluate and report in} Fig.~\ref{fig:fluctuations} for the three cases in presence of the EF and against the fluctuations computed in liquid water without EF (cyan circles). For all cases, we determine $\sigma(n)$ in time windows of $50$~ps. In the case of $0.05$~V/{\AA}, we can observe that, with respect to the case in the absence of the field, the fluctuations are strongly d\textcolor{black}{a}mped but for hexagonal and pentagonal rings, which fluctuate in a comparable measure. Upon increasing the simulation time, the fluctuations of the HBN are reduced for all cases but for the hexagonal rings, which become increasingly enhanced with the simulation time. Considering that the sample is liquid (although with strongly reduced diffusion), we posit that the diffusion occurs via changes in the HBN mostly involving hexagonal rings. \newline
At $0.10$~V/{\AA} and $0.15$~V/{\AA}, we observe a drastic suppression of the fluctuations of the HBN, \textcolor{black}{to values well below those of the liquid}. Such marked reduction of the fluctuations \textcolor{black}{is responsible for the suppression of long-range density fluctuations occurring in correspondence with the transition to glassy water~\cite{martelli2022steady,formanek2023molecular}, a characteristic that differentiates liquid water from glassy states~\cite{martelli2017large}. Therefore, our findings are strongly indicative of a transition to a} glass.

\section{Discussion}\label{sec3}
In this work, we have performed long \emph{ab initio} simulations of bulk water at \emph{ambient conditions} in the presence of applied external electric fields (EFs) in the range $0.05\leq$EFs$\leq0.15$~V/{\AA}. We have inspected the out-of-equilibrium process at disjoint time windows and recorded the results on each window. In the presence of an EF of $0.05$~V/{\AA}, the dipoles align along the direction parallel to the EF while the diffusivity becomes sluggish. Overall, the inspected quantities computed within the last $\sim150$~ps of simulation are stable in time, indicating that the system is genuinely a liquid. \newline 
Upon increasing the EF to $0.10$~V/{\AA} and to $0.15$~V/{\AA}, we observe a transition to a new ferroelectric glass that we call f-GW (ferroelectric glassy water). The amorphization occurs after $\sim150-200$~ps and is signaled by the freezing of the translational degrees of freedom and a drop in the potential energy, indicating that the sample has reached a metastable basin on the potential energy landscape. The evolution in time of the radial distribution functions \textcolor{black}{and of other structural descriptors} report an enhancement of the first and the second shells of neighbours along with a drastic depletion of the entries populating the space between them, as expected in the low-density glassy water. Similarly, the hydrogen bond network (HBN) undergoes a progressive structural reorganization favoring hexagonal motifs and a corresponding suppression of its fluctuations, as expected in the low-density glassy water state~\cite{martelli2022steady}.

Our work represents the first evidence of electrofreezing of liquid water at \emph{ambient conditions}, a task that has been attempted since 1862~\cite{Dufour_1862}. The new f-GW phase \textcolor{black}{can be unveiled only by accessing and isolating late portions of long AIMD simulations, and} is a new tile in the complex phase diagram of water. Therefore, it enriches our understanding of the physics of this complex material. Nonetheless, the conditions explored in this work are ubiquitous in industrial and natural settings, fields that can potentially benefit from this work. For example, water is routinely exposed to natural EFs comparable to the ones explored in this work when at the interface with enzymes, proteins, and biological membranes, defining the biological functionality and stabilizing such complex structures. 

We infer that an experimental validation of our finding and the realization of the f-GW phase might be relatively straightforward by exploiting modern experimental settings. Many laboratories are nowadays capable of quantifying the field strengths generated in the proximity of emitter tips~\cite{aragones,che,balke} -- such those established by STM and AFM apparatus --, which fall in the same range required to transition to the new f-GW. Nonetheless, we posit that lower fields may induce electrofreezing to f-GW on longer time scales, \textcolor{black}{accessible to accurate interaction potentials such as, e.g., MB-Pol~\cite{MB-pol2023} or Neural Network potentials~\cite{zhang2018deep}}. 


\section{Methods}\label{sec4}
\subsection{Numerical simulations}
We performed \emph{ab initio} molecular dynamics (AIMD) simulations using the software package CP2K~\cite{CP2K}, based on the Born-Oppenheimer approach. The external electric fields (EFs) are static, homogeneous and directional (i.e., along the $z$-axis). The implementation of external EFs in Density Functional Theory (DFT) codes can be achieved \emph{via} the modern theory of polarization and Berry's phases~\cite{MPT1,MPT2,Berry}. In particular, owing to the seminal work carried out by Umari and Pasquarello~\cite{Umari}, nowadays AIMD simulations under the effect of static EFs with periodic boundary conditions are routinely performed. The reader who is interested in the implementation of EFs in atomistic simulations can refer to the following literature: Refs.~\cite{Umari,MPT1,MPT2,Nunes1,Nunes2,Resta,Gonze1,Gonze2}. The \textcolor{black}{main} simulation here presented consists of a liquid water sample containing $128$ H$_{2}$O molecules arranged in a cubic cell with side parameter $a=15.82$~{\AA}, so as to reproduce a density of $0.97$~g$\cdot$cm$^{-3}$. Furthermore, additional simulations were executed on bigger cubic cells composed of $256$ water molecules and having edges of $20.05$~{\AA} and $20.26$~{\AA}. In such a case, lower densities of $0.95$~g$\cdot$cm$^{-3}$ and $0.92$~g$\cdot$cm$^{-3}$ were simulated, respectively. To minimize \textcolor{black}{undesirable} surface effects, the structures were replicated in space by employing periodic boundary conditions. We applied static and homogeneous EFs of intensities equal to $0.05$~V/{\AA}, $0.10$~V/{\AA}, and $0.15$~V/{\AA} from a zero-field condition in parallel simulation runs. The maximum field strength of $0.15$~V/{\AA} was chosen to prevent water splitting known to occur at larger field intensities~\cite{Stuve2012,Hammadi2012,LeeNanoRes,Saitta_PRL,Cassone_JPCL}. In the zero-field case we performed dynamics of $50$~ps whereas, for each other value of the field intensity, we ran dynamics of at least $250$~ps. Besides, as for the simulations of the lower-density states only a single field intensity of $0.15$~V/{\AA} was simulated -- in addition to the fieldless cases -- for time-scales of $\sim500$~ps ($\rho=0.95$~g$\cdot$cm$^{-3}$) and $\sim450$~ps ($\rho=0.92$~g$\cdot$cm$^{-3}$). This way, we accumulated a global simulation time approaching $2$~ns, whilst a time-step of $0.5$~fs has been chosen. 

Wavefunctions of the atomic species have been expanded in the TZVP basis set with Goedecker-Teter-Hutter pseudopotentials using the GPW method~\cite{GTH3}. A plane-wave cutoff of $400$~Ry has been imposed. Exchange and correlation (XC) effects were treated with the gradient-corrected Becke-Lee-Yang-Parr (BLYP)~\cite{BLYP1,BLYP2} density functional. Moreover, in order to take into account dispersion interactions, we employed the dispersion-corrected version of BLYP (i.e., BLYP+D3(BJ))~\cite{Grimme1,Grimme2}. The adoption of the BLYP+D3 functional has been dictated by the widespread evidence that such a functional, when dispersion corrections are taken into account, offers one of the best adherence with the experimental results among the standard GGA functionals~\cite{Lin_JCTC}. It is well-known, indeed, that neglecting dispersion corrections leads to a severely over-structured liquid (see, e.g., Ref.~\cite{Gillan} and references therein). Moreover, a nominal temperature slightly higher than the standard one has been simulated in the main simulations to better reproduce the liquid structure (i.e., $T=350$~K). Furthermore, the additional simulations at lower density regimes were executed at a lower (supercooling) temperature of $T=250$~K (see the SI for the respective results).  

Albeit the BLYP+D3 functional represents e reasonably good choice, computationally more expensive hybrid functionals, such as revPBE0, when simulated along with the quantum treatment of the nuclei performs excellently well for water, as demonstrated by Marsalek and Markland~\cite{marsalek}. 
However, since sufficiently large simulation boxes are necessary to track structural transitions, the inclusion of the nuclear quantum effects is beyond the scope of the present work. Moreover, IR absorption line shapes of liquid water (and ice) are overall reproduced remarkably well by standard AIMD simulations, which include by their nature the explicit quantum adiabatic response of the electrons~\cite{sharma}. In addition, the adherence of the IR and of the Raman spectra evaluated by some of us~\cite{Cassone_PCCP19} under zero-field conditions with recent experimental results~\cite{Bertie,pattenaude} justifies \emph{a posteriori} the classical treatment of the nuclei.   
As a consequence, the dynamics of ions was simulated classically within a constant number, volume, and temperature (NVT) ensemble, using the Verlet algorithm whereas the canonical sampling has been executed by employing a canonical-sampling-through-velocity-rescaling thermostat~\cite{Bussi} set with a time constant equal to $10$~fs. IR spectra have been determined by means of the software TRAVIS~\cite{Travis} \textcolor{black}{(see the SI for further information)}.  

\subsection{Network topology}
In order to probe the topology of the hydrogen bond network (HBN), we employed ring statistics, a theoretical tool that has proven to be instrumental in investigating the network topology in numerically simulated network-forming materials. The ring statistics is only one of many graph-based techniques to investigate network topologies and, in the case of water, it has helped in understanding the connections between water anomalies and thermodynamic response functions~\cite{martelli2019unravelling,formanek2020probing} as well as the properties of glassy water~\cite{formanek2023molecular}. We construct rings by starting from a tagged water molecule and recursively traversing the HBN until the starting point is reached or the path exceeds the maximal ring size considered ($10$ water molecules in our case). The definition of hydrogen bond follows Ref.~\cite{luzar1996hydrogen}. We do not distinguish between the donor-acceptor character of the starting water molecule.




\backmatter

\bmhead{Supplementary information}

A Supporting Information (SI) file with additional analyses and results accompanies the current work. 

\bmhead{Acknowledgments}

G.~C. acknowledges support from ICSC – Centro Nazionale di Ricerca in High Performance Computing, Big Data and Quantum Computing, funded by European Union – NextGenerationEU - PNRR, Missione 4 Componente 2 Investimento 1.4. G.~C. is thankful to CINECA for an award under the ISCRA initiative, for the availability of
high performance computing resources and support.





\bigskip

\bibliography{sn-bibliography}

\end{document}


\title[]{Supporting Information -- Electrofreezing of Liquid Water at Ambient Conditions}


\author*[1]{\fnm{Giuseppe} \sur{Cassone}}\email{cassone@ipcf.cnr.it}

\author*[2,3]{\fnm{Fausto} \sur{Martelli}}\email{fausto.martelli@ibm.com}


\affil[1]{\orgdiv{Institute for Chemical-Physical Processes}, \orgname{National Research Council}, \orgaddress{\street{Viale F. Stagno d'Alcontres 37}, \city{Messina}, \postcode{98158}, \country{Italy}}}

\affil[2]{\orgname{IBM Research Europe}, \orgaddress{\street{Keckwik Lane}, \city{Daresbury}, \postcode{WA4 4AD}, \country{United Kingdom}}}

\affil[3]{\orgdiv{Department of Chemical Engineering}, \orgname{University of Manchester}, \orgaddress{\street{Oxford Road}, \city{Manchester}, \postcode{M13 9PL}, \country{United Kingdom}}}



\maketitle

\section{Additional results}\label{sec2}

\textcolor{black}{Infrared spectra shown in Fig.~1 of the main text have been determined by means of the software TRAVIS~\cite{Travis1,Travis2} from the centers of the Maximally Localised Wannier Functions (MLWFs)~\cite{MLWF1,MLWF2} calculated on the fly during the \emph{ab initio} molecular dynamics (AIMD) simulations.
Molecular dipoles from MLWFs centers can be determined as:
\begin{equation}
    \mu=-2e\sum_{i}\textbf{r}_{i}+e\sum_{j}Z_{j}\textbf{R}_{j} \, ,
\end{equation}
where $e$ is the electron charge, $\textbf{r}_{i}$ is the position vector of the MLWF center $i$, $Z_j$ is the atomic number of the nuclei $j$ whilst $\textbf{R}_{j}$ is the position vector of this latter. This way, the IR spectra at the investigated field intensities were computed as the Fourier transform of the molecular dipole autocorrelation function along the last $50$ ps of the respective simulation trajectories.
}

In order to track molecular reorientations under the field action, we compute the distributions of the angle $\theta$ formed between the instantaneous water molecular dipole vectors and the field direction (i.e., $z$-axis), Fig.~S1.
\begin{figure}[h!]
\centering
    \includegraphics[width=0.8\textwidth]{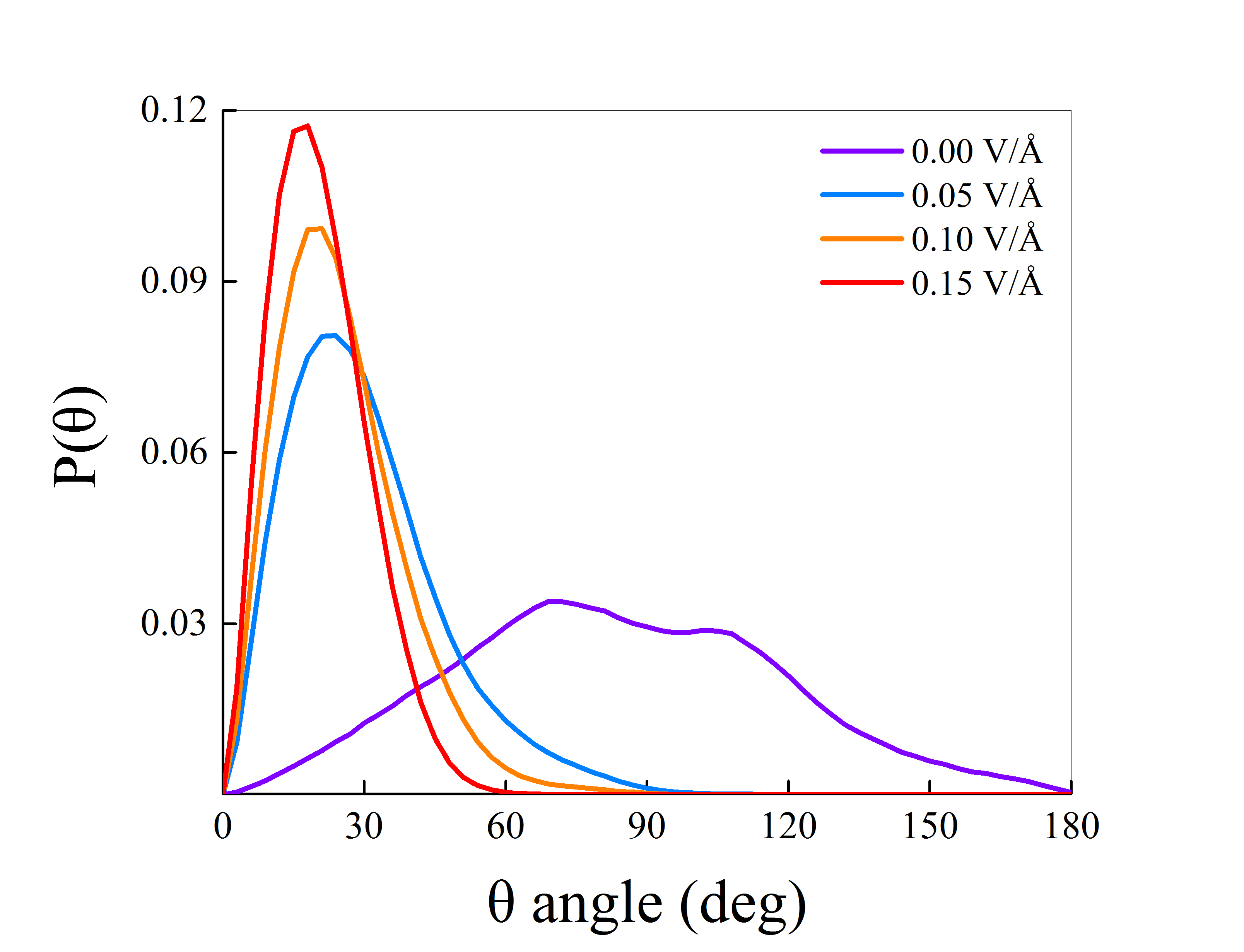}
    \caption{Distributions of the $\theta$ angle formed between the instantaneous molecular dipole vectors during the last $50$~ps of the respective trajectories and the EF direction for diverse field intensities applied along the $z$-axis.}
    \label{fig:lsi}
\end{figure}
Interestingly, whilst the field is capable of reorienting a large fraction of water dipoles already at $0.05$~V/{\AA}, the electrostatic potential gradient producing this field strength does not induce a net suppression of the translational degrees of freedom of the molecules, as shown in Fig.~3 of the main text. The enhancement of the water dipoles at increasingly high fields is also visible from the dipole distributions reported in Fig.~S2-a, showing a progressive shift towards larger magnitudes and a slight narrowing of the distributions.
\begin{figure}[h!]
    \centering
    \includegraphics[width=\textwidth]{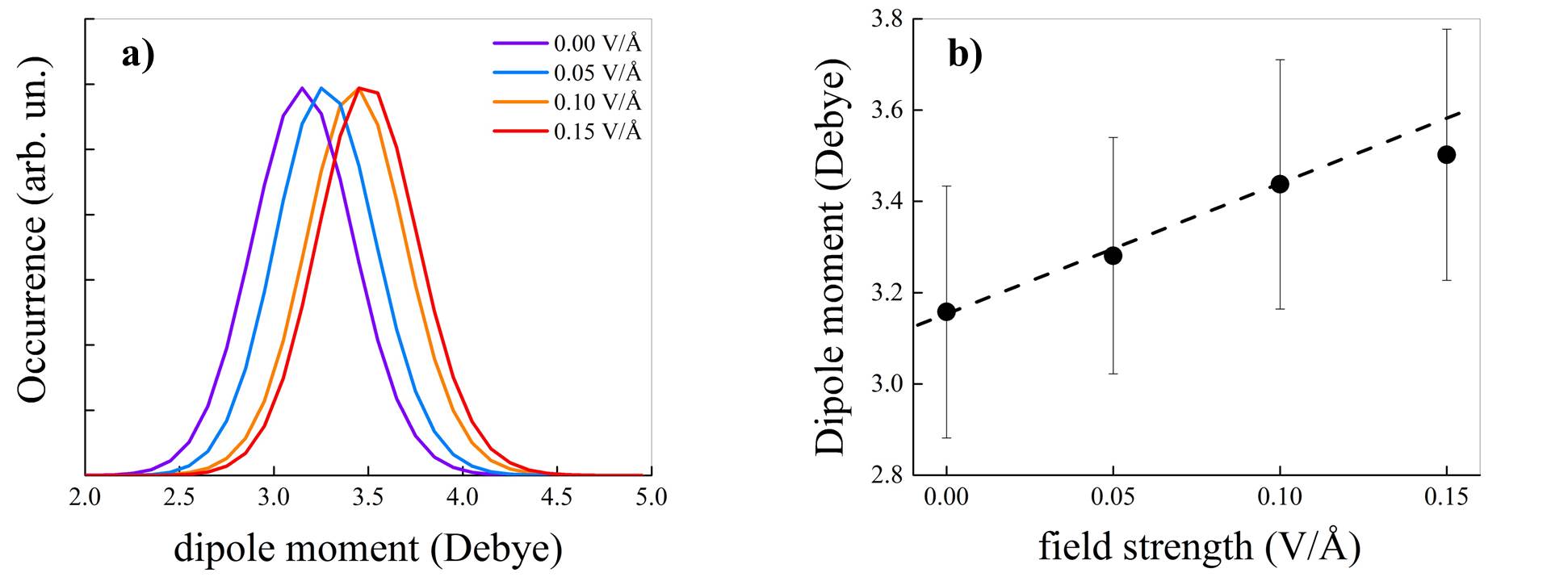}
    \caption{(a) Distributions of the magnitude of the water dipoles extracted from the last $50$~ps of the respective simulations at different field intensities \textcolor{black}{determined from the MLWFs centers}. (b) Average water dipole \textcolor{black}{and associated standard deviation} extracted from the distributions in (a). It is noteworthy pointing out the interruption of the linear response regime for field strengths producing water electrofreezing.}
    \label{fig:fig1}
\end{figure}
\begin{figure}[h!]
    \centering
    \includegraphics[width=\textwidth]{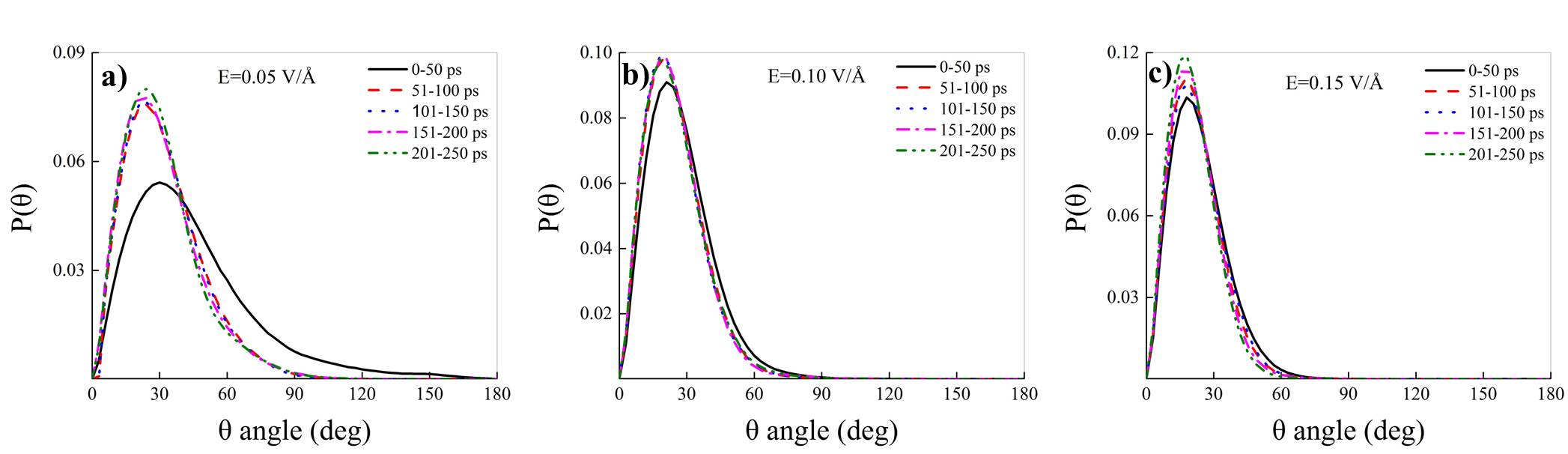}
    \caption{\textcolor{black}{Distributions of the $\theta$ angle at $0.05$~V/{\AA} (a), $0.10$~V/{\AA} (b), and $0.15$~V/{\AA} (c) measured at consecutive time windows. The black solid curves refer to the first $50$~ps, dashed red lines to the time window $51-100$~ps, dotted blue curves to the time window $101-150$~ps, the dashed-dotted magenta lines to the time window $151-200$~ps, and the dashed-dotted-dotted green curves to the window $201-250$~ps.}}
    \label{fig:fig1}
\end{figure}

\begin{figure}[h!]
    \centering
    \includegraphics[width=\textwidth]{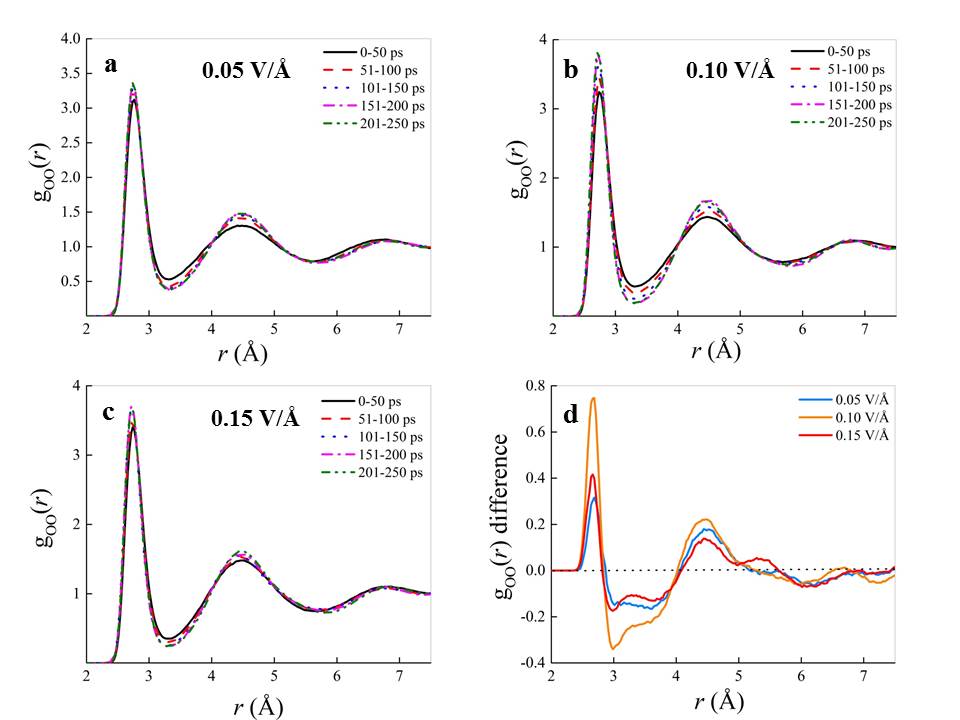}
    \caption{Oxygen-oxygen radial distribution functions determined for disjoint time frames of $50$~ps each determined from the \emph{ab initio} molecular dynamics simulations conducted at $0.05$~V/{\AA} (a), $0.10$~V/{\AA} (b), and $0.15$~V/{\AA} (c) field strengths. The black solid curves refer to the first $50$~ps, dashed red lines to the time window $51-100$~ps, dotted blue curves to the time window $101-150$~ps, the dashed-dotted magenta lines to the time window $151-200$~ps, and the dashed-dotted-dotted green curves to the window $201-250$~ps. In panel (d), the point-by-point difference of the oxygen-oxygen radial distribution functions determined in the final and the initial time frames (i.e., during the last $50$ and first $50$~ps, respectively) for diverse field intensities is shown.}
    \label{fig:fig1}
\end{figure}
\begin{figure}[h!]
    \centering
    \includegraphics[width=0.8\textwidth]{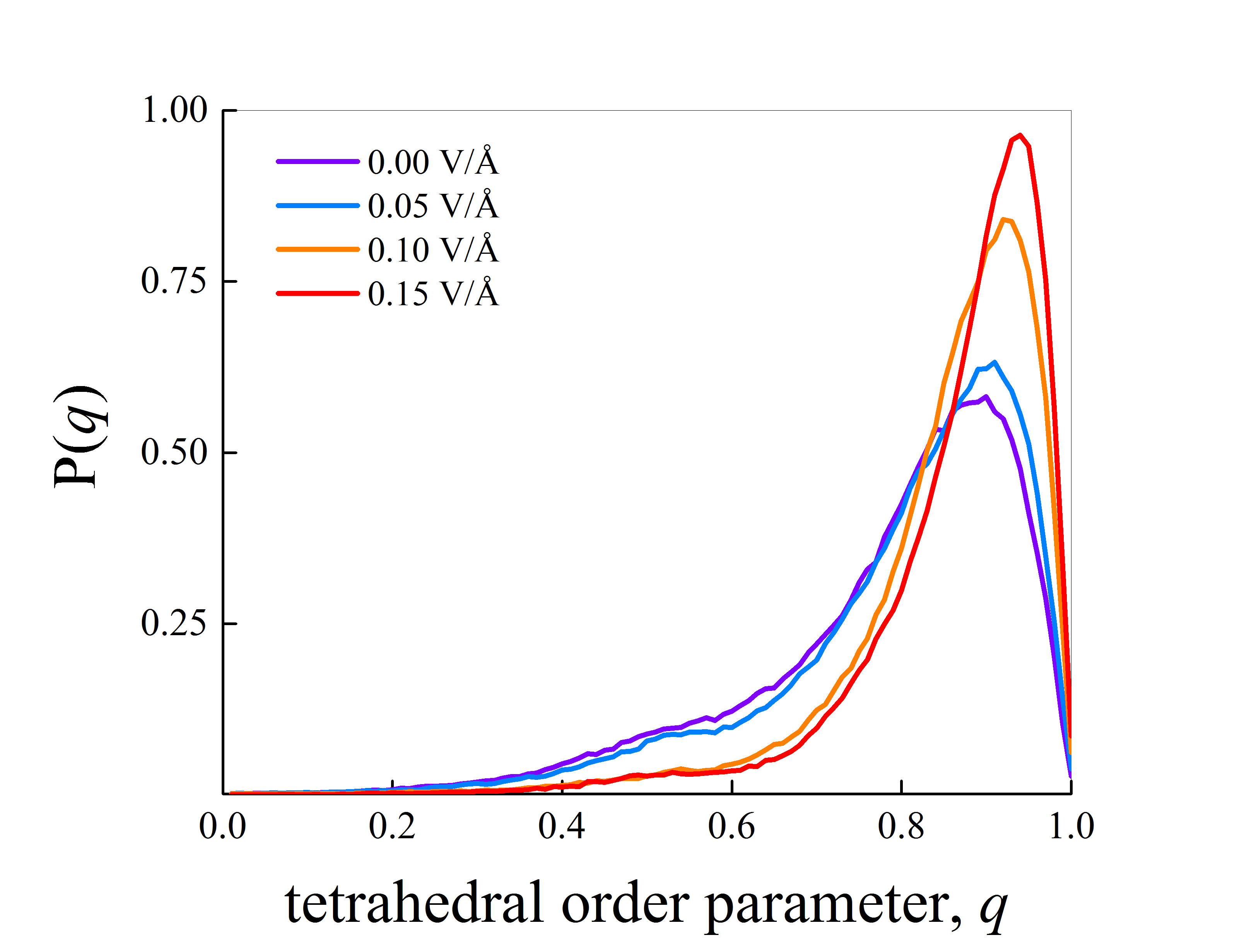}
    \caption{Distribution of the orientational tetrahedral order parameter $q$ determined at zero field (violet curve) and at field strengths equal to $0.05$~V/{\AA} (blue curve), $0.10$~V/{\AA} (orange curve), and $0.15$~V/{\AA} (red curve).}
    \label{fig:fig1}
\end{figure}
Fig.~S2-b reports the profile of the dipole moment with the field strength. It is possible to recognize a linear regime holding up to a strength of $0.10$~V/{\AA}. Thus, the transition from the liquid to the f-GW phase is also marked by the breakdown of the linear response regime to external electric fields (EFs).
\textcolor{black}{Additionally to this analysis, it is worth monitoring the temporal dependence of the P($\theta$) distributions at disjoint time windows, a procedure allowing for disclosing the dynamical response of the sample. As displayed in Fig.~S3, the field-induced reorientation of the molecular dipoles takes place on fast timescales and achieves saturation within the first $50$~ps of dynamics at all field intensities, with the exception of the weakest field (Fig. S3-a), where nonetheless the convergence of the dipolar response is reached in less than $100$~ps. }

Fig.~S\textcolor{black}{4} reports the oxygen-oxygen radial distribution functions computed at consecutive, disjoint time windows of $50$~ps. At $0.05$~V/{\AA}, the $g_{OO}(r)$ converges to a steady profile after $50$~ps (Fig.~S\textcolor{black}{4}-a), while convergence is achieved only after $150$~ps for $0.10$~V/{\AA} (Fig.~S\textcolor{black}{4}-b) and $200$~ps for $0.15$~V/{\AA} (Fig.~S\textcolor{black}{4}-c). The $g_{OO}(r)$ computed within the last $50-100$~ps for $0.10$~V/{\AA} and $0.15$~V/{\AA} resemble the $g_{OO}(r)$ of a low-density amorphous (LDA) (see main text). To shed some light on the dynamical reorganization of the water structure induced by the external field, we have evaluated the oxygen-oxygen radial distribution function differences between the last $50$~ps and the first $50$~ps time frames of each simulation, as reported in Fig.~S\textcolor{black}{4}-d. Whereas at $0.05$~V/{\AA} structural differences between the initial and the final time windows appear to be small -- as also visible in Fig.~S\textcolor{black}{4}-a --, a field of intensity equal to $0.10$~V/{\AA} induces much larger global reorganizations towards more structured molecular correlations in the system (Fig.~S\textcolor{black}{4}-d, yellow curve). On the other hand, the evidence that these differences are smaller in the sample exposed to a $0.15$~V/{\AA} field (Fig.~S\textcolor{black}{4}-d, red curve) has to be ascribed to a faster initial reorganization taking place since the first $50$~ps of dynamics, whereas longer timescales ($\sim200$~ps) are somehow needed for bringing to completion the structural transition in the simulated sample, as shown in Fig.~S\textcolor{black}{4}-c.

In Fig.~S\textcolor{black}{5} we report $P(q)$, the distribution of the tetrahedral order parameter $q$ defined as
\begin{equation}
q=1-\frac{3}{8}\sum_{j=1}^3\sum_{k=j+1}^4\left(\cos \psi_{jk}+\frac{1}{3} \right)^2
\end{equation}
where $\psi_{jk}$ is the angle formed between the oxygen atoms of the water molecule under consideration and its nearest neighbour oxygen atoms $j$ and $k$. The tetrahedral order parameter $q$ was originally proposed by Chau and Hardwick~\cite{chau1998new} and subsequently rescaled by Errington and Debenedetti~\cite{Errington} so that the average value of $q$ varies from $0$ for an ideal gas to $1$ for a regular tetrahedron. From Fig.~S\textcolor{black}{5} it is possible to observe that the samples in the absence of a field and in the presence of a field of $0.05$~V/{\AA} show a very similar tetrahedral character. A major change occurs at stronger fields signaling the transition to the more ordered f-GW phase.

Similar conclusion\textcolor{black}{s} can be drawn upon inspecting the local structure index (LSI)~\cite{shiratani1996growth,shiratani1998molecular}, an insightful order parameter that can be employed to characterize the LDL and HDL molecular environments and defined as the inhomogeneity on the distribution of radial distances
\begin{equation}
I=\frac{1}{N}\sum_{j=1}^N\left[\Delta_{j+1,j}-\left<\Delta\right> \right]^2
\end{equation}
where $\Delta_{j+1,j}=r_{j+1}-r_j$ is the distance between particles within a cutoff distance of $3.7$~{\AA} from a reference molecule and $\left<\Delta\right>$ is the average overall neighbours of a molecule within the given cutoff. The LSI, therefore, provides a convenient quantitative measure of the fluctuations in the distance distribution surrounding a given water molecule within a sphere defined by a radius of $3.7$~{\AA}. In doing so, the index $I$ measures the extent to which a given water molecule is surrounded by well-defined first and second coordination shells. In Fig.~S\textcolor{black}{6}, we report the LSI computed for the three EFs here inspected at time windows of $50$~ps. It is possible to observe the development of hints of a bimodal distribution in the cases of $0.10$~V/{\AA} and $0.15$~V/{\AA} in correspondence with the transition to f-GW. \textcolor{black}{This can be also appreciated from the lower panel of Fig.S6, reporting the LSI computed in the last time window $[201-250]$~ps for $0.15$V/{\AA} and for the LDA simulated via classical molecular dynamics at $T=200$~K. The latter has been obtained upon quenching liquid water from $T=300$~K to $T=200$~K at a quenching rate of $1$~K/ns, as reported in Refs.~\cite{martelli2017large,martelli2022steady,martelli2018searching,formanek2023molecular}}
\begin{figure}[h!]
    \includegraphics[width=\textwidth]{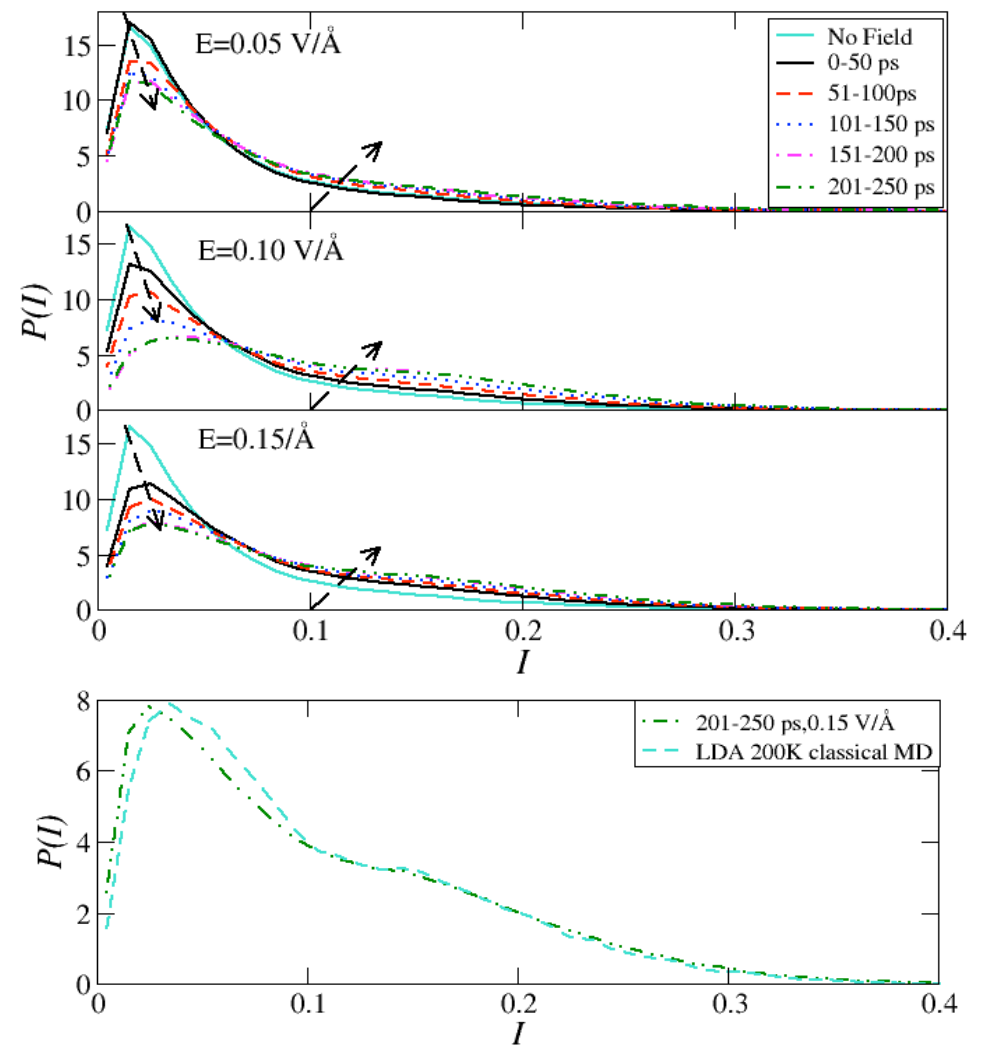}
    \caption{Upper panel: $P(I)$ computed for the case at $0.05$~V/{\AA}, $0.10$~V/{\AA}, and $0.15$~V/{\AA} at disjoint time windows of $50$~ps. \textcolor{black}{Lower panel: comparison between the LSI computed for $0.15$V/{\AA} and the LSI computed for LDA at $T=200$~K from classical molecular dynamics.}}
    \label{fig:lsi}
\end{figure}

\textcolor{black}{Somewhat related to the local and global degree of order of the H-bond network, is its kinetics. In particular, we performed a structural analysis of the H-bond network and identified a H-bond through the following geometric conditions (that must be simultaneously fulfilled): two water molecules are considered as H-bonded if $R^{(OO)}\leq3.5$~{\AA} and $\angle$O-H$\cdot\cdot\cdot$O$\leq30^{\circ}$, where $R^{OO}$ is the instantaneous distance between the oxygen atoms. From this, we calculated the time autocorrelation function of H-bonds as:
\begin{equation}
c(t)=\frac{\sum_{\langle{i,j}\rangle}{s_{ij}(t_{0})s_{ij}(t_{0}+t)}}{\sum_{\langle{i,j}\rangle}{s_{ij}(t_{0})}} \, ,
\end{equation}
where the indices $i$ and $j$ run on all pairs of first-neighbour molecules which at $t_{0}$ were H-bonded, $t_{0}$ being the time at which the measurement process begins; $s_{ij}=1$ if the criterion for the presence of a H-bond is fulfilled, $s_{ij}=0$ otherwise. The results were averaged over hundreds of initial configurations. Fig.~S7 shows the continuous (Fig.~S7-a) and intermittent (Fig.~S7-b) autocorrelation functions $c(t)$ of the H-bonds for different field intensities. Within the intermittent definition of $c(t)$, a given H-bond is allowed to cleave within timescales $\leq5$~fs to account for bond fluctuations. Thus, within this latter fast timescale, we always assign to $s_{ij}$ a value equal to $1$ when considering the intermittent autocorrelation function (see eq.~(4)).}
\begin{figure}[h!]
    \includegraphics[width=\textwidth]{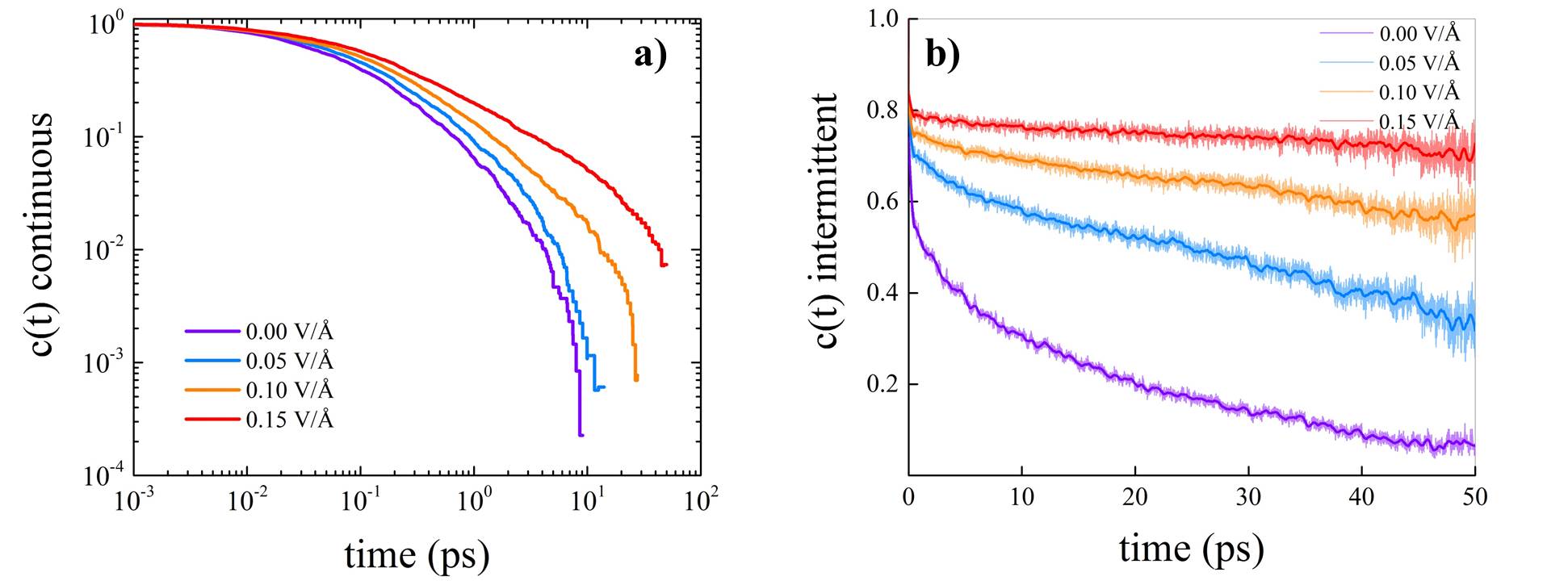}
    \caption{\textcolor{black}{Log-log continuous (a) and linear-scale intermittent (b) H-bond autocorrelation functions calculated during the last $50$~ps of the respective simulations.}}
    \label{fig:lsi}
\end{figure}
\textcolor{black}{The application of a $0.05$~V/{\AA} field induces only a relatively moderate -- with respect to the zero-field case -- slow down of the dynamics of the H-bond network recorded by means of the continuous $c(t)$ function (Fig.~S7-a). Instead, significantly more drastic effects are recorded upon applying fields of $0.10$ and $0.15$~V/{\AA}. Interestingly, the changes produced on the H-bond network kinetics by these field regimes qualitatively resemble those induced by a sizable ($\sim40$~K) decrease of the temperature~\cite{Stanley_PRE}. This is also visible from the intermittent H-bond autocorrelation function displayed in Fig.~S7-b. Although the H-bond characteristic time recorded at zero field (violet curve) is extended by the application of a field strength of $0.05$~V/{\AA} (blue curve), a visible decay of the intermolecular correlations within the timescales of our simulations is recorded at the latter regime. Instead, a field of $0.10$~V/{\AA} (orange curve) and a field of $0.15$~V/{\AA} (red curve) clearly strengthen the H-bond persistence over sizably longer timescales. These results are fully consistent with the picture emerging from the partial Van Hove correlation functions shown in the main text (Fig.~2).  }

In Fig.~S\textcolor{black}{8} we report the oxygen-oxygen radial distribution function computed with a larger simulation box of $256$ water molecules, at densities of $0.92$~g$\cdot$cm$^{-3}$ and $0.95$~g$\cdot$cm$^{-3}$ and at a temperature of $250$~K (panels (a) and (b), respectively) for a sample without EF and a sample with a field of $0.15$~V/{\AA}. These simulations reach $\sim500$~ps. It is possible to observe the development of an f-GW-like $g_{OO}(r)$ at both densities, indicating that the transition to f-GW reported in our work is not an artifact of small simulation boxes \textcolor{black}{and that takes place for different densities}.
\begin{figure}[h!]
    \centering
    \includegraphics[width=\textwidth]{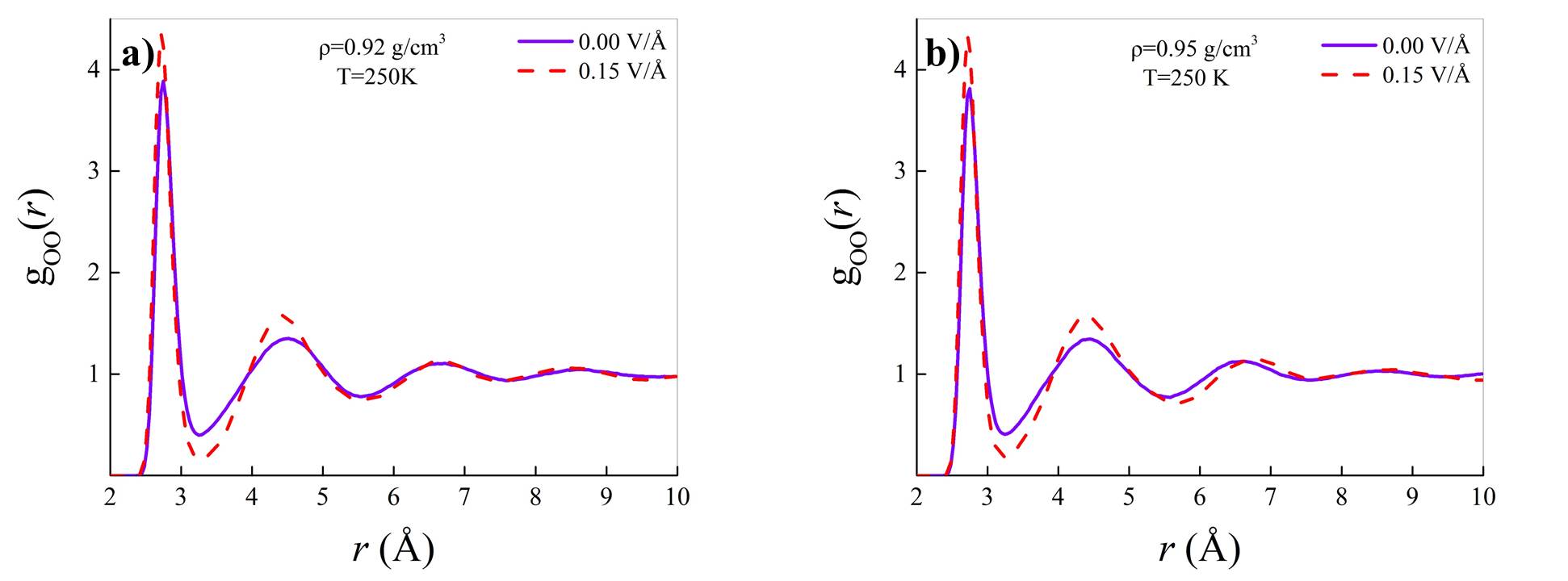}
    \caption{Oxygen-oxygen radial distribution functions in the absence of the field (violet solid lines) and in the presence of a field strength equal to $0.15$~V/{\AA} from \emph{ab initio} molecular dynamics simulations conducted on boxes containing $256$ H$_2$O molecules in the supercooled regime ($T=250$~K), and for densities of $0.92$~g$\cdot$cm$^{-3}$ (a) and $0.95$~g$\cdot$cm$^{-3}$ (b).}
    \label{fig:fig1}
\end{figure}

\begin{figure}[h!]
    \includegraphics[width=\textwidth]{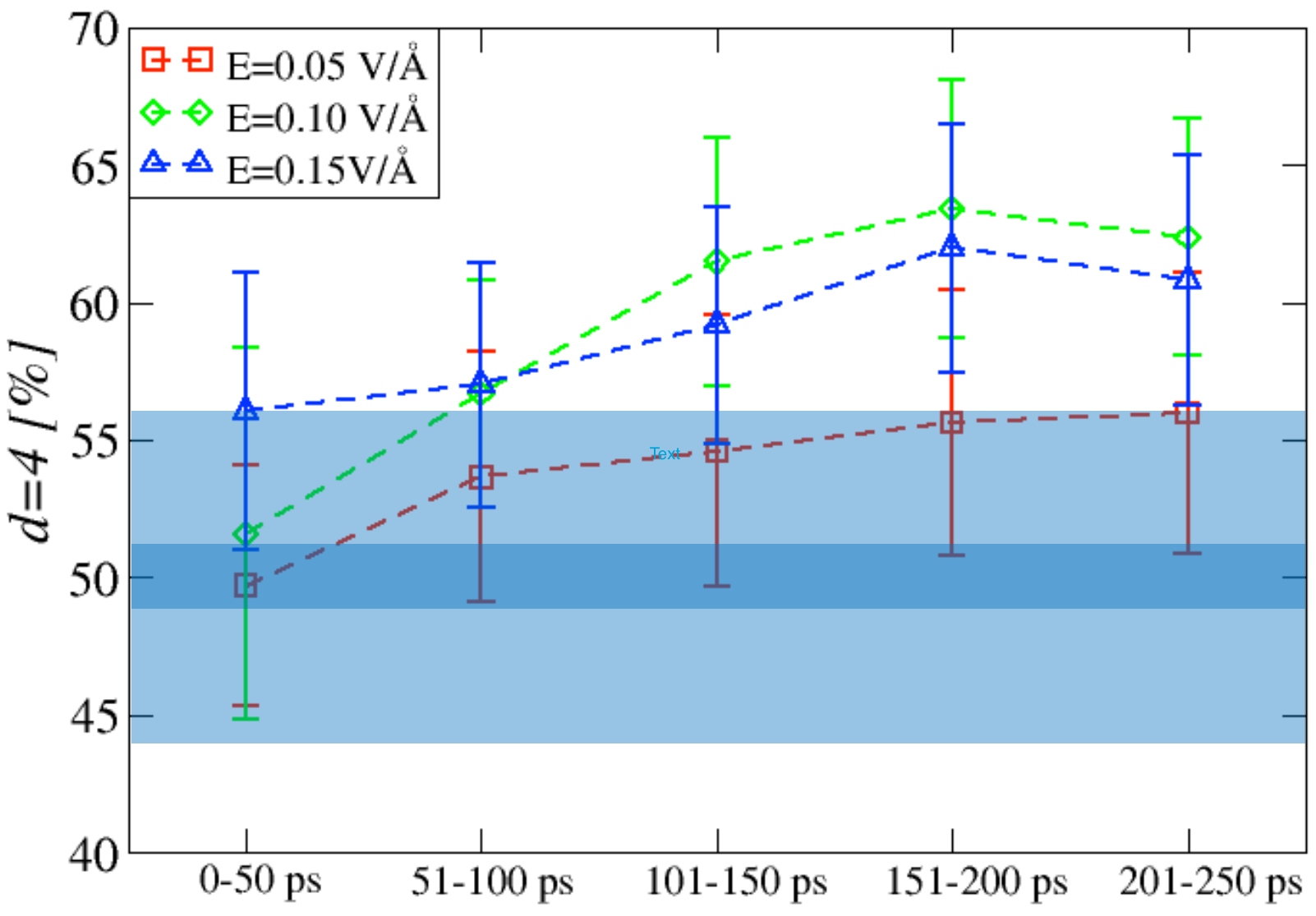}
    \caption{Percentage of four-coordinated water molecules computed on consecutive time windows and compared against the case of bulk water in the absence of EFs (blue stripe). Red circles refer to the case of $0.05$~~V/{\AA}, green diamonds to $0.10$~V/{\AA}, blue triangles to $0.15$~V/{\AA}.}
    \label{fig:HBtime}
\end{figure}

\textcolor{black}{In Fig.~\ref{fig:HBtime} we report $d=4$, the percentage of four-folded water molecules at consecutive time windows. The blue stripe corresponds to the case of liquid water in the absence of EFs. In the presence of $0.05$~V/{\AA} (red squares), $d=4$ increases from $\sim50\%$ to $\sim53\%$ within the first $50$~ps of the simulation, and keeps gradually increasing reaching a maximum of $\sim56\%$ in the last two time windows. Upon increasing the field to $0.10$~V/{\AA} (green diamonds) we can observe that $d=4$ computed within the first $50$~ps is roughly the same as the case for $0.05$~V/{\AA} computed on the same time window. On the other hand, the $d=4$ linearly increases by $\sim6\%$ in the second and in the third time window. The $d=4$ reaches then a plateau in correspondence with the last two time windows. Upon increasing the field strength to $0.15$~V/{\AA}, we observe a sudden increase in the $d=4$ to $\sim56\%$ within the first time window. Further increases occur at the later stages of the simulation as for the cases previously inspected with lower field strengths. It is worth noticing that the percentage of four-coordinated water molecules for $0.10$~V/{\AA} and $0.15$~V/{\AA} is almost indistinguishable towards the end of the simulation.}

\begin{figure}[h!]
    \centering
    \includegraphics[width=\textwidth]{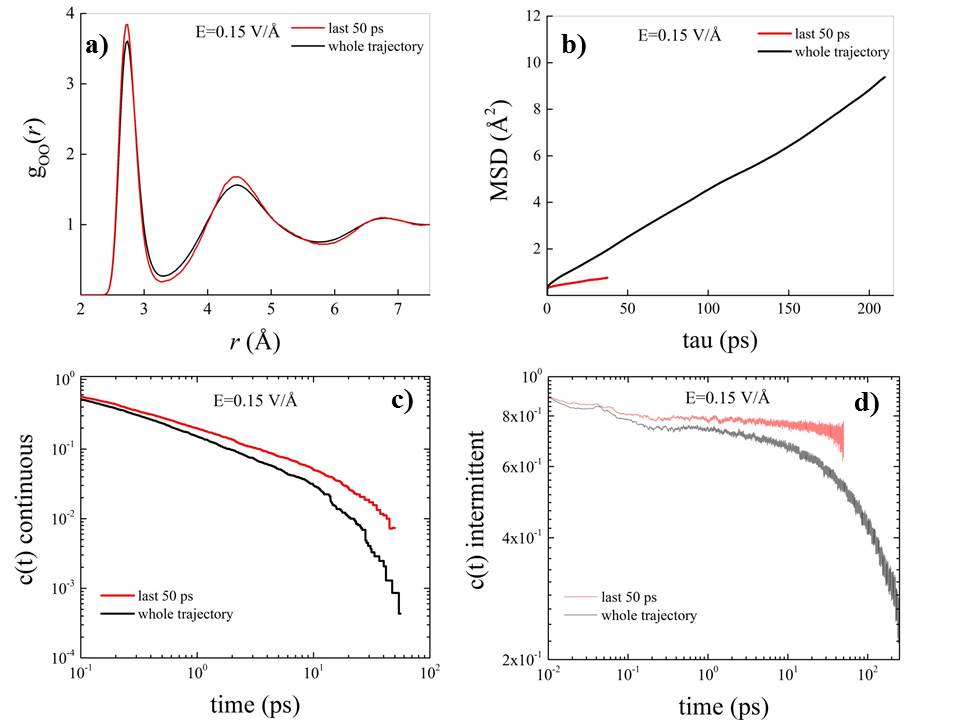}
    \caption{\textcolor{black}{(a) Oxygen-oxygen radial distribution functions, (b) oxygen mean squared displacement, (c) log-log continuous and (d) intermittent autocorrelation functions of the H-bonds determined at an EF intensity of $0.15$~V/{\AA} calculated over the whole trajectory (black lines) and during the last $50$~ps of dynamics (red lines).}}
    \label{fig:fig1}
\end{figure}
\textcolor{black}{To further stress the relevance of the sampling at disjoint time windows, we report in Fig.~S10 a series of structural and dynamical observables determined at the highest field intensity here explored (i.e., $0.15$~V/{\AA}) and calculated over the whole $250$-ps-long trajectory and on the last $50$~ps of dynamics. The $g_{OO}(r)$ at $0.15$~V/{\AA} determined over the whole trajectory exhibits smaller (higher) peaks (dips) with respect to the same quantity calculated over the last $50$~ps of dynamics of the same trajectory, as displayed in Fig.~S10-a. Interestingly, the importance of sampling at consecutive time frames is even more visible from dynamical rather than structural properties. In fact, to adequately evaluate the field effects on the translational degrees of freedom, the sampling at consecutive time windows here adopted appears to be necessary for the timescales affordable by \emph{ab initio} simulations. As shown in Fig.~S10-b, indeed, the mean squared displacement of the oxygen atoms determined over the whole trajectory at $0.15$~V/{\AA} witnesses the mixing of different translational regimes, a circumstance leading to an underestimation of the EF-induced damping effect.
This is not only true for translational but also for rotational degrees of freedom, which are intimately related to the dynamics of the H-bond network. By direct comparison of the continuous (Fig.~S10-c) and intermittent (Fig.~S10-d) H-bond autocorrelation functions determined over different timescales (see legends), diverse H-bond characteristic times emerge. All these findings prove that relevant information on the effects produced by the application of external fields on liquid water can be unveiled only by accessing and isolating late portions of long \emph{ab initio} simulations. Exclusively by adopting this strategy it is possible to catch the \emph{electrofreezing} effect induced by the field on the roto-translational degrees of freedom of water and, presumably, of other H-bonded systems.}

\newpage
\newpage

\backmatter

\bibliography{sn-bibliography}